\documentclass[aps,prd,twocolumn,showpacs,10pt,superscriptaddress,preprintnumbers,nofootinbib]{revtex4-1}
\usepackage{epsfig,amssymb,amsmath,psfrag,epstopdf,color}
\usepackage{graphicx}
\usepackage{caption}
\usepackage{subcaption}
\usepackage{paralist}

\allowdisplaybreaks

\def\beq{\begin{equation}}
\def\eeq{\end{equation}}
\def\bsp#1\esp{\begin{split}#1\end{split}}
\newcommand{\be}{\begin{equation}}
\newcommand{\ee}{\end{equation}}
\newcommand{\bea}{\begin{eqnarray}}
\newcommand{\eea}{\end{eqnarray}}
\newcommand{\eqn}[1]{eq.~\eqref{#1}}

\def\eqn#1{eq.~(\ref{#1})}
\def\eqns#1#2{eqs.~(\ref{#1}) and~(\ref{#2})}

\def\to{\rightarrow}

\def\ksl{\not{\hbox{\kern-2.3pt $k$}}}

\def\e{\epsilon}

\def\Ord{{\cal O}}

\def\Neqfour{{{\cal N}=4}}
\def\Neqone{{{\cal N}=1}}

\def\eqn#1{eq.~(\ref{#1})}
\def\eqns#1#2{eqs.~(\ref{#1}) and~(\ref{#2})}

\def\spa#1.#2{\left\langle#1\,#2\right\rangle}
\def\spb#1.#2{\left[#1\,#2\right]}
\def\lor#1.#2{\left(#1\,#2\right)}
\def\sand#1.#2.#3{%
\left\langle\smash{#1}{\vphantom1}^{-}\right|{#2}%
\left|\smash{#3}{\vphantom1}^{-}\right\rangle}

\newcommand{\dd}{\mathrm{d}}
\newcommand{\nn}{\nonumber}
\newcommand{\brk}{\nonumber\\&}
\newcommand{\nbrk}{\nonumber\\}

\newcommand{\gcusp}{\Gamma^{\mathrm{cusp}}}
\newcommand{\gsoft}{\gamma^{s}}

\def\im{\mathrm{i}}

\def\sss{\scriptscriptstyle}
\def\qt{q_{\sss T}}
\def\vecqt{\vec{q}_{\sss T}}
\def\als{\alpha_{\sss S}}
\def\as{a_{\sss S}}
\def\scetii{SCET$_{\sss \textrm{II}}$ }
\def\nuw{\nu}
\def\gae{\gamma_{\sss E}}
\def\sfd{S_{\rm \sss F.D.}}
\def\esfd{S^{\rm \sss F.D.}}
\def\gc{\Gamma_{\mathrm{cusp}}}
\def\gs{\gamma_s}

\def\gr{\gamma_{r}}
\def\ugr{\gamma^{r}}
\def\bals#1{\big[\als(#1)\big]}
\def\vecb{\vec{b}_\perp}
\def\LQ{L_{\sss Q}}

\begin{document}

\preprint{FERMILAB-PUB-16-067-PPD-T}
\preprint{MIT-CTP-4796}

\title{Bootstrapping rapidity anomalous dimension for transverse-momentum resummation}
\author{Ye~Li}
\email{yli32@fnal.gov}
\affiliation{Fermilab, PO Box 500, Batavia, IL 60510, USA}
\author{Hua~Xing~Zhu}
\email{zhuhx@mit.edu}
\affiliation{Center for Theoretical Physics, Massachusetts
  Institute of Technology,
Cambridge, MA 02139, USA}

\begin{abstract}
\noindent Soft function relevant for transverse-momentum resummation
for Drell-Yan or Higgs production at hadron colliders are computed
through to three loops in the expansion of strong coupling, with the
help of bootstrap technique and supersymmetric decomposition. The
corresponding rapidity anomalous dimension is extracted. An intriguing
relation between anomalous dimensions for transverse-momentum
resummation and threshold resummation is found. 
\end{abstract}

\maketitle
 
\noindent\textbf{Introduction.} The
transverse-momentum~($\qt$) distribution of generic high-mass
color-neutral systems~(Drell-Yan lepton pair, Higgs, EW vector boson
pair, etc.) produced in hadron collisions is of great
interest since the early days of Quantum
ChromoDynamics~(QCD)~\cite{PHLTA.B79.269,NUPHA.B154.427,NUPHA.B193.381,NUPHA.B197.446,NUPHA.B250.199,Arnold:1990yk,Ladinsky:1993zn,Ellis:1997ii,Balazs:1997xd,Qiu:2000ga,hep-ph/0008184,Berger:2002ut,Bozzi:2005wk,Rogers:2010dm,Aybat:2011zv,Becher:2011xn,Su:2014wpa}. It
provides a testing ground for examination and improvement of our
understanding of QCD, both perturbatively and non-perturbatively. 
When $\qt$ is 
small compared with the invariant mass $Q$ of the system, fixed-order perturbation
theory breaks down due to the appearance of large logarithms 
of the form $\ln^k (\qt^2/Q^2)/\qt^2$, with $k\geq 0$ at each order in
strong coupling $\als$. These large logarithms originate from incomplete
cancellation of soft and collinear divergences between real and
virtual diagrams.  Fortunately, Collins, Soper, and Sterman~(CSS) have
shown that they
can be systematically resummed to all orders in perturbation
theory~\cite{NUPHA.B250.199}, thanks to QCD factorization. 

In recent years, there have been increasing interests in applying
Soft-Collinear Effective
Theory~(SCET)~\cite{hep-ph/0005275,hep-ph/0011336,hep-ph/0109045,hep-ph/0202088,hep-ph/0206152}
to resum large logarithms in perturbative QCD using 
renormalization group~(RG) method. For $\qt$ resummation this has been
done by a number of
authors~\cite{hep-ph/0501229,hep-ph/0507196,0911.4135,1007.4005,1104.0881,1202.0814,1111.4996}. For
transverse-momentum observable, the relevant momentum modes in
light-cone coordinate
for fields in the effective theory are soft $p_s \sim Q(\lambda,  
\lambda, \lambda)$, collinear $p_c\sim Q(\lambda^2, 1, \lambda)$ and
anti-collinear $p_{\bar{c}} \sim Q (1, \lambda^2 ,\lambda)$.
Here $\lambda\sim \qt/Q$ is a power counting parameter. The corresponding
effective theory is \scetii. An important feature of \scetii  is that
soft and collinear modes live on the same hyperbola of
virtuality,
$p^2_s \sim p^2_c \sim p^2_{\bar c} \sim \lambda^2 Q^2$. Besides the usual
large logarithms of ratio between hard scale $Q$ and soft scale
$\lambda Q$, there are also large rapidity separations between soft,
collinear, and anti-collinear modes which need to be resummed. 
In
this Letter we adopt the rapidity RG formalism of Chiu, Jain, Neill,
and Rothstein~\cite{1104.0881,1202.0814}. According to the rapidity RG
formalism, cross section at small $\qt$ factorizes into hard function $H$,
Transverse-Momentum-Dependent~(TMD) beam functions $B$, and TMD soft
function $S_{\perp}$. Schematically the factorization formula reads:
\begin{align}
  \frac{1}{\sigma}\frac{d^3\sigma^{(\mathrm{res.})}}{ d^2\vecqt\,dY\,dQ^2}
  \sim & \,H(\mu) \int \frac{d^2\vec{b}_\perp}{(2\pi)^2}\, e^{\im \vec{b}_\perp \cdot \vec{Q}_T}
\nbrk
&\, \cdot    [B\otimes B](\vecb,\mu,\nuw) S_{\perp}(\vec{b}_\perp,\mu,\nuw)  
\label{eq:scetfac}
\end{align}
Large logarithms in virtuality is
resummed by running in the renormalization scale $\mu$, while large logarithms in rapidity is
resummed by running in the rapidity scale $\nuw$. The $\mu$ evolution of the hard
function can be derived from quark or gluon form factor and is
well-known~\cite{0902.3519,1001.2887,1004.3653}. Since the physical
cross section is independent of $\mu$ and $\nuw$ order by order in
the perturbation theory, it follows that the $\mu$ and $\nuw$ evolution of
$[B\otimes B]$ is fixed once the corresponding evolution for the soft
function is known. The knowledge of $\mu$ and $\nuw$ evolution of
hard, beam, and soft function, together with the boundary conditions of
these functions at initial scales, determine the all order structure of large
logarithms of $\qt$. 

The naive definition of the TMD soft function is a vacuum expectation value of
light-like Wilson loops with a transverse separation, which suffers from light-cone/rapidity
divergence~\cite{NUPHA.B193.381}. A proper definition of the TMD soft function requires the
introduction of appropriate regulator for the rapidity divergence.
 Proposals to regularize the rapidity
divergence includes non-light-like axial gauge without Wilson
lines~\cite{NUPHA.B250.199}, tilting Wilson lines off the lightcone~\cite{hep-ph/0404183},
nearly light-like Wilson lines with subtraction of soft
factor~\cite{Collins:2011zzd}, modifying the phase space measure~\cite{1007.4005,1104.0881,1112.3907}, modifying
the $\im \varepsilon$ prescription of eikonal propagator~\cite{1208.1281},
etc. In this Letter, we follow the recent proposal~\cite{Li:2016axz} by Neill and the
current authors of implementing an infinitesimal shift in the time
direction to the Wilson loop correlator. Specifically, the TMD soft function
with the rapidity regulator of Ref.~\cite{Li:2016axz} reads:  
\begin{align}
  \label{eq:1}
  S_{\perp} (\vecb,\mu,\nuw) = & \,\lim_{\nuw \to + \infty} S_{\rm \sss F.D.} (\vecb,\mu,\nuw)
\\
\equiv &\, \lim_{\nuw \to +\infty}  \frac{1}{d_a}\big\langle 0 \big|
  \mathrm{T}\big[S^\dagger_{\bar{n}}(-\infty,0)S_n(0,-\infty)\big]
\nbrk
& \,
\cdot
\overline{\mathrm{T}}\big[S^\dagger_{n}(-\infty,y_\nuw(\vec{b}_\perp))S_{\bar
  n}( y_\nuw(\vec{b}_\perp),-\infty)\big]   \big| 0 \big\rangle
\nonumber
\end{align}
where the two Wilson loops are separated by the distance $y_\nuw(\vec{b}_\perp) =  ( \im \, b_0/ \nuw,\, \im \, b_0/ \nuw, \, \vec{b}_\perp)$, with $b_0 = 2 e^{-\gae}$.
$S_{n(\bar{n})}$ are path-ordered Wilson lines on the light-cone. They
carry fundamental or adjoint color indices, depending on whether the
color-neutral system is produced in $q\bar{q}$ annihilation ($d_a =
N_c$) or $gg$ fusion ($d_a = N^2_c - 1$). $\mathrm{T}$ is the time-ordered
operator. The soft function $S_{\perp}$ in \eqn{eq:1} is closely
related to the so-called fully differential soft function~\cite{0911.4135},
$\sfd$. The limit $\nuw\to +\infty$ means that only the non-vanishing terms
of $\sfd$ are kept in that limit. The important role of $\sfd$ in our calculation will be explained in
the next section. Note that our definition for the TMD soft function doesn't
rely on perturbation theory. However, we restrict to the
perturbatively calculable part of the soft function in this Letter.
 
After minimal subtraction of dimensional regularization pole $1/\e^n$
in $\overline{\textrm{MS}}$ scheme,
the soft function $S_{\perp}$ depends on both the renormalization scale
$\mu$ and the rapidity scale $\nuw$. The $\mu$ evolution of the TMD soft function is
specified by the RG equation:
\begin{align}
  \label{eq:3}
  \frac{\dd \ln S_{\perp}(\vecb,\mu,\nuw) }{\dd \ln \mu^2} = \gc\bals{\mu} \ln\frac{\mu^2}{\nuw^2} - \gs\bals{\mu}
\end{align}
where $\gc$ is the well-known
light-like cusp anomalous dimension~\cite{NUPHA.B164.171,PHLTA.B171.459}, which
is known to three loops in QCD~\cite{hep-ph/0403192}. $\gs$ is the soft 
anomalous dimension governing the single logarithmic evolution, which
can be extracted through to three loops from QCD splitting
function~\cite{hep-ph/0403192} and quark and gluon form
factor~\cite{0902.3519,1001.2887,1004.3653}, as is confirmed by
explicit three-loop calculation~\cite{1412.2771}. The
rapidity evolution equation for the TMD soft function reads: 
\begin{align}
  \label{eq:4}
  \frac{\dd \ln S_{\perp}(\vecb,\mu,\nuw)}{\dd \ln \nuw^2} =&\,
  \int^{b^2_0/\vec{b}_\perp^{\, 2}}_{\mu^2}\! \frac{\dd \bar{\mu}^2}{\bar{\mu}^2}
  \gc\bals{\bar{\mu}}  \brk
+
\gr\bals{b_0/|\vec{b}_\perp|}
\end{align}
where the rapidity anomalous dimension $\gr$ is introduced for the single
logarithmic evolution of rapidity logarithms. Thanks to the non-Abelian
exponentiation
theorem~\cite{Sterman:1981jc,PHLTA.B133.90,NUPHA.B246.231} which our
regularization procedure~\cite{Li:2016axz} preserves, the perturbative soft function can be
written as an exponential:
\begin{align}
  \label{eq:5}
\!\!\!  S_{\perp}(\vecb,\mu,\nu) = \exp\Big[ \as S^{\perp}_1  + \as^2 S^{\perp}_2 + \as^3 S^{\perp}_3 +
  \Ord(\as^4)\Big]
\end{align}
where we have defined $\as = \als(\mu)/(4\pi)$ as our perturbative
expansion parameter throughout this Letter. The one and two-loop
coefficients $S^{\perp}_{1,2}$ can be found in Ref.~\cite{Li:2016axz}. In the
next section we outline the procedure we used to calculate the
three-loop coefficient $S^{\perp}_3$, from
which the rapidity anomalous dimensions can be extracted to
the same order. 
 

\noindent\textbf{Method.} To obtain the TMD soft function $S_{\perp}$ through to
three loops, we first calculate the fully differential soft function
to the same order. $\sfd$ obeys a RG equation identical to
\eqn{eq:3}~\cite{0911.4135}:
\begin{align}
  \label{eq:6}
  \frac{\dd \ln \sfd(\vecb,\mu,\nuw) }{\dd \ln \mu^2} = \gc\bals{\mu} \ln\frac{\mu^2}{\nuw^2}  - \gs\bals{\mu}  
\end{align}
In $\sfd$, $\nuw$ is a parameter
of the theory, not a regulator. Therefore the $\nuw$ dependence of $\sfd$ is in general complicated. The perturbative solution
to $\sfd$ is then determined by \eqn{eq:6} and the boundary condition
at initial scale, $\sfd(\vecb,\mu=\nuw,\nuw)$. Similar to $S_{\perp}$, $\sfd$ can also be
written as an exponential, as in \eqn{eq:5}. The one and two-loop
coefficients $\esfd_{1,2}$ were first computed in Ref.~\cite{1105.5171}, and
reproduced in Ref.~\cite{Li:2016axz}.

By dimension analysis, $\sfd(\vecb,\nu,\nu)$ is a function of $x = -
\vecb^{\, 2}\nuw^2/b^2_0$. A strategy based on the bootstrap program for
scattering amplitudes~\cite{Dixon:2011pw} is proposed in
Ref.~\cite{Li:2016axz} to compute $\sfd(\vecb,\nu,\nu)$, which we briefly recall
below. In Ref.~\cite{1105.5171}, the one and two-loop coefficients
$\esfd_{1,2}$ are written in terms of classical and Nielsen’s
polylogarithms with argument $x$. 
A crucial observation made in Ref.~\cite{Li:2016axz} is that
the same results can be written in terms of harmonic polylogarithms~(HPL)
$H_{\vec{w}}(x)$~\cite{hep-ph/9905237},  with weight indices drawn
from the set $\{0,1\}$. Furthermore, for the available one and
two-loop data, the leftmost and the rightmost index of the weight vectors were found to be $0$ and $1$,
respectively. The rightmost index has to be $1$, because the two cusp points of the Wilson loops are separated by Euclidean distance for $x<0$, and no branch cut is
expected. On the other hand, the condition on the leftmost-index comes empirically from the observation of the one- and two-loop results; as we will show below, this condition breaks down at three loops in QCD. Nevertheless, for now we proceed with the empirical ansatz
for $L$-loop fully differential soft function proposed in Ref.~\cite{Li:2016axz}, which
is a linear combination of HPLs with undetermined rational
coefficients, and whose weight vectors obey the leftmost- and rightmost-index conditions. 
The undetermined coefficients of the HPLs can then be
fixed by performing an expansion around $x\sim 0$, together with the
constraint that rapidity divergence is only a single logarithmic
divergence at each order for the expansion coefficients in \eqn{eq:5}.
It turns out that the $x \to 0$ limit of $\sfd$ is smooth, and
the expansion is simply a Taylor series in $x$. As explained in
Ref.~\cite{Li:2016axz}, the leading $x^0$ term of the expansion reproduces
the threshold soft function~\cite{1412.2771}, while the coefficient of
$x^n$ can be obtained by inserting a numerator $(l^+l^- - l^2)^n$ into the
integrand of the threshold soft function, where $l$ is the total
momenta of real radiation from the time-ordered Wilson loop. Furthermore, using
Integration-By-Parts~(IBP)
identities~\cite{NUPHA.B192.159,hep-ph/0102033}, integrals with high rank numerator insertion can be reduced
to a small number of master integrals, which have been computed for
other purpose recently~\cite{1302.4379,1309.4391,1309.4393,1404.5839,1501.00236,1505.04110}.

Although the strategy
outlined above is straightforward, it has two caveats. First, the
maximal weight of HPLs at three loops for massless perturbation theory
is $6$. It follows that the number of coefficients need to be fixed is
$\sum^4_{i=0} 2^i = 31$. In other words, one needs to insert a high-rank numerator $(l^+l^- -
l^2)^{31}$ into the integrand of threshold soft function in order to
have enough data to fix the coefficients, which is unfortunately
beyond  the ability of the 
tools for IBP
reduction~\cite{hep-ph/0404258,0807.3243,1201.4330,1212.2685}. Second,
it is not clear whether the conjectured sets of function in
Ref.~\cite{Li:2016axz} is sufficient to describe the three-loop soft
function. To circumvent the above difficulties, we first perform the
calculation for soft Wilson loops whose matter content~\cite{1309.4391,1404.5839,1412.2771} resembles those of
$\Neqfour$ Supersymmetric Yang-Mills theory~(SYM). This has a
number of advantages:
\begin{inparaenum}[1)]
\item it has been observed that for soft
Wilson loops in SCET~\cite{1412.2771}, the results in $\Neqfour$ SYM has uniform
degrees of transcendentality with transcendental weight $2L$ at $L$
loops. Furthermore, the $\Neqfour$ results match the maximal-weight
part of the corresponding QCD results. Similar phenomenon was first
observed for anomalous dimension of twist-two operator for Wilson
lines~\cite{hep-ph/0301021}. It also holds for some other quantities,
e.g., perturbative form factor~\cite{0902.3519,1112.4524,Brandhuber:2012vm}. Assuming that this is also true in our
current calculation, by calculating $\sfd$ in $\Neqfour$ SYM first, we
should automatically obtain the maximal-weight part of $\sfd$ in
QCD;
\item since the $\Neqfour$ SYM results have uniform degrees of
  transcendentality, there are only $16$ coefficients to be fixed at
  three loops, which can be achieved within the current computation
  power;
\item the remaining parts of the QCD result have transcendental weight
  lower than $6$, therefore only requires $15$ coefficients to be
  fixed. Alternatively, since the Feynman diagrams corresponding to
  the lower-weight part have less complicated analytical structure,
  they can be computed by brute force. Direct calculation can also
  test the completeness of the ansatz. And it turns out that although the
  ansatz remain complete for the three-loop $\Neqfour$ SYM result, it
  fails for the three-loop QCD one. Fortunately, for QCD result, a
  brute-force calculation for the terms proportional to $n_f$ is
  possible using the method of Ref.~\cite{1501.00236}. More
  importantly, the result for $n_f$ terms indicates which set of
  functions we should add to the existing ansatz. 
\end{inparaenum}
The full results,
for both $\Neqfour$ SYM and QCD, are presented in the next section.


\noindent\textbf{Results.} We first present the results for $\sfd$ in $\Neqfour$
SYM. We only give the results at the initial scale, $\mu=\nuw$. The
full scale dependence can be inferred from \eqn{eq:6}. The one and two-loop coefficients can be found in
Ref.~\cite{Li:2016axz}. The three-loop coefficient in the
four-dimensional-helicity scheme~\cite{hep-ph/0202271} reads
\begin{widetext}
\vspace{-1.5em}
\begin{align}
\esfd_{3,\sss\Neqfour}\Big|_{\mu = \nuw} = & c^s_{3,\sss\Neqfour}  +  N_c^3 \Big(16 \zeta_2 H_{4}+48 \zeta_2 H_{2,2}+64
  \zeta_2 H_{3,1}
+96 \zeta_2 H_{2,1,1}+120 \zeta_4 H_{2}+48 H_{6}+24 H_{2,4}
+40 H_{3,3}
\brk
+72 H_{4,2}
+128 H_{5,1}+16 H_{2,1,3}
+56 H_{2,2,2}+80 H_{2,3,1}+80 H_{3,1,2}+144 H_{3,2,1}
+224 H_{4,1,1}
\brk
+64 H_{2,1,1,2}+96 H_{2,1,2,1}
+160 H_{2,2,1,1}+256 H_{3,1,1,1}+192 H_{2,1,1,1,1} \Big)
\label{eq:neq4n3lo}
\end{align}
\end{widetext}
\vspace{-3em}
where $c^s_{3,\sss\Neqfour}= 492.609 N^3_c$ is the three-loop constant for threshold
soft function in $\Neqfour$ SYM~\cite{1412.2771}. We have used the
shorthand notation for the HPLs~\cite{hep-ph/9905237} and neglected
the argument $x$. It is
interesting to note that each term in \eqn{eq:neq4n3lo}
has uniform sign and integer coefficient. Furthermore, overall sign is alternating at each
order in $\als$~\cite{Li:2016axz}. Similar behavior of alternating
uniform signs in perturbative expansion with increasing loop order for
certain observable was
known before, see Ref.~\cite{Henn:2012qz}.
The corresponding results for QCD in 't Hooft-Veltman scheme reads:
\begin{widetext}
\vspace{-2em}
  \begin{align}
    \label{eq:7}
    \esfd_{3}\Big|_{\mu=\nuw} =&\, c^s_3 + \frac{C_a C_A^2}{N^3_c}
                                 \left(\esfd_{3,\sss\Neqfour}(x)\Big|_{\mu
                                 = \nuw} - c^s_{3,\sss\Neqfour}\right)
  + C_a C_A^2 \Biggl[ -\frac{1072}{9} \zeta_2 H_{2}-176 \zeta_3
                                 H_{2}-\frac{88}{3} \zeta_2 H_{3}+88
                                 \zeta_2 H_{2,1}
\brk
+\frac{30790}{81}H_{2}+\frac{7120}{27} H_{3}-\frac{104}{9} H_{4}-\frac{440}{3}
                                 H_{5}-\frac{8}{3}
                            \left(H_{1,1}-\frac{H_{1,1}}{x}\right)-\frac{7120}{27}
                                 H_{2,1}-\frac{1072}{9}
                                 H_{2,2}-\frac{88}{3}
                                 H_{2,3}
\brk
-\frac{3112}{9} H_{3,1}
-88 H_{3,2}-\frac{352}{3} H_{4,1}-\frac{392}{3} H_{2,1,1}+\frac{88}{3}
                                 H_{2,1,2}+\frac{352}{3}
                                 H_{2,2,1}+\frac{352}{3} H_{3,1,1}+352
                                 H_{2,1,1,1} \Biggr]
\brk
+ C_a C_A n_f \Biggl[ \frac{160}{9} \zeta_2 H_{2}+\frac{16}{3} \zeta_2
                                 H_{3}-16 \zeta_2
                                 H_{2,1}-\frac{7988}{81}
                                 H_{2}-\frac{2312}{27}
                                 H_{3}-\frac{64}{3} H_{4}+\frac{80}{3}
                                 H_{5}+\frac{8}{3}
                                 \left(H_{1,1}-\frac{H_{1,1}}{x}\right)
\brk
+\frac{2312}{27} H_{2,1}+\frac{160}{9} H_{2,2}+\frac{16}{3}
                                 H_{2,3}+\frac{224}{3} H_{3,1}+16
                                 H_{3,2}+\frac{64}{3}
                                 H_{4,1}-\frac{32}{9}
                                 H_{2,1,1}-\frac{16}{3}
                                 H_{2,1,2}-\frac{64}{3} H_{2,2,1}
\brk
-\frac{64}{3} H_{3,1,1}-64 H_{2,1,1,1} \Biggr]
+ C_a n_f^2 \Biggl(\frac{400}{81} H_{2}+\frac{160}{27}
H_{3}+\frac{32}{9} H_{4}-\frac{160}{27} H_{2,1}
-\frac{32}{9} H_{3,1}+\frac{32}{9} H_{2,1,1} \Biggr)
\brk
+  C_a C_F n_f \Biggl( 32 \zeta_3 H_{2}-\frac{110}{3}
  H_{2}-8 H_{3}+8 H_{2,1} \Biggr)
  \end{align}
\end{widetext}
where $C_a=C_F$ for Drell-Yan process, and $C_a = C_A$ for Higgs
production. $c^s_3$ is the three-loop scale independent part of the
treshold soft function in QCD, $c^3_s = S^{\rm thr.}_3(\tau,\mu =
\tau^{-1})$, see for example Refs.~\cite{1412.2771,Anastasiou:2016cez,Li:2016axz}. It can be found in eq.~(3.2) of Ref.~\cite{1412.2771}
by multiplying a casimir rescaling factor $C_a/C_A$. We note that the
only term that goes beyond the empirical ansatz~\cite{Li:2016axz} is
$(H_{1,1} - H_{1,1}/x)$~\footnote{This term cancels out in the
  $\Neqfour$ combination, as is clear from \eqn{eq:neq4n3lo}. It also
  cancels out in the pure $\Neqone$ SYM with adjoint gluino, in which one simply sets $n_f
\to C_A$ and $C_F \to C_A$. We thank Mingxing Luo and Lance Dixon for
pointing out this.}, which can be inferred from the direct calculation of
the $n_f$-dependent part using Feynman diagram method. Specifically, if all the relevant
integrals are known, the result for $\Neqfour$ SYM in
\eqn{eq:neq4n3lo} can also be
obtained using Feynman diagram method, in a gauge theory
with $n_f=4$ adjoint fermions, $n_s=6$ adjoint real scalars, and with
proper Yukawa interaction between the fermions and scalars. While the
integrals for the pure gluon contribution are challenging, we
manage to compute the $n_f$- and $n_s$-dependent terms by brute-force Feynman
diagram calculation. We observe that for both the fermion and scalar contributions, the only addition needed to correct the empirical ansatz at three loops is the combination $(H_{1,1} - H_{1,1}/x)$. From
there we can readily extract the gluon contribution, which is the
same in $\Neqfour$ SYM and QCD, by subtracting from \eqn{eq:neq4n3lo}
the corresponding fermion and scalar contributions. We can also conclude that the only addition to the ansatz of the gluon contribution is the combination $(H_{1,1} - H_{1,1}/x)$.  

We briefly describe the available checks on our results in
\eqns{eq:neq4n3lo}{eq:7}. Firstly, as mentioned above, due to the relative
simplicity in the resulting integrals, we have been able to compute all the
$n_f$-dependent part in \eqn{eq:7} by directly calculating the
Feynman diagrams. We find that our ansatz, even including the
$(1-1/x)H_{1,1}$ term, is insufficient to express the result in
the intermediate step of the direct calculation. The additional terms
needed are $(1-1/x)H_{1}$, $H_{2}/x$, $\zeta_2 H_1 -
H_{1,2}$. Interestingly, they all cancel out in the sum of real and
virtual contributions. Secondly, our ansatz can be uniquely
fixed at three loops using the data from Taylor expansion over $x$
through to $x^{10}$. However, we have obtained the expansion data
through to $x^{17}$, leading to an over constrained system of
equations. We found that the solution exist and is unique for the
system, thus providing a strong check of our calculation. See,
e.g. Ref.~\cite{Henn:2013wfa} for similar discussion on using over constrained
system of equations to fix ansatz.

With the fully differential soft function at hand, it is
straightforward to obtain $S_{\perp}$ by taking the limit $\nuw \to +
\infty$ using the package \texttt{HPL}~\cite{hep-ph/0507152}. The soft anomalous dimension $\gs$ through to three loops can be
found, e.g., in eq.~(A.4-6) of Ref.~\cite{1412.2771} by an rescaling
factor $C_a/C_A$. The rapidity anomalous dimensions are given by:
\begin{align}
  \label{eq:8}
  \ugr_0 = & \, 0
\nbrk
\ugr_1 = &\, C_a C_A  \left(28 \zeta_3-\frac{808}{27}\right)+\frac{112
  C_a n_f}{27}
\nbrk
\ugr_2 = &\, C_a C_A^2 \Biggl( -\frac{176}{3} \zeta_3
           \zeta_2+\frac{6392 \zeta_2}{81}+\frac{12328
           \zeta_3}{27}+\frac{154 \zeta_4}{3}
\brk
-192\zeta_5-\frac{297029}{729} \Biggr) + C_a C_A n_f \Biggl(
           -\frac{824 \zeta_2}{81}-\frac{904 \zeta_3}{27}
\brk
+\frac{20\zeta_4}{3}+\frac{62626}{729}
\Biggr) + C_a n_f^2 \Biggl(-\frac{32 \zeta_3}{9}-\frac{1856}{729} 
\Biggr)
\brk
+ C_a C_F n_f \Biggl( -\frac{304
           \zeta_3}{9} -16 \zeta_4+\frac{1711}{27} \Biggr)
\end{align} 
Note that $\ugr_0$ and $\ugr_1$ can be obtained from QCD anomalous
dimension known long time
ago~\cite{Davies:1984hs,NUPHA.B256.413,hep-ph/0008152}. They have also been
reproduced in SCET recently~\cite{Gehrmann:2014yya,1511.05590,1602.01829,Li:2016axz}. The
three-loop coefficient $\ugr_2$  is new and is one of the main results of this Letter.
It is also straightforward to obtain the boundary condition of $S_{\perp}$
at the initial scale, $c^{\perp}_3 \equiv S^{\perp}_3(\vecb,\mu=b_0/|\vecb|,\nuw=b_0/|\vecb|)$:
\begin{align}
  \label{eq:9}
  c^{\perp}_3 = &\, C_a C_A^2 \Biggl( \frac{928 \zeta_3^2}{9}+\frac{1100}{9}
            \zeta_2 \zeta_3-\frac{151132 \zeta_3}{243}-\frac{297481
            \zeta_2}{729}
\brk
+\frac{3649 \zeta_4}{27}+\frac{1804
            \zeta_5}{9}-\frac{3086 \zeta_6}{27}+\frac{5211949}{13122}
\Biggr)
\brk
+ C_a C_A n_f \Biggl(\frac{40}{9} \zeta_3 \zeta_2+\frac{74530
            \zeta_2}{729}+\frac{8152 \zeta_3}{81}-\frac{416
            \zeta_4}{27}
\brk
-\frac{184 \zeta_5}{3}-\frac{412765}{6561} \Biggr)
+ C_a C_F n_f \Biggl( -\frac{80}{3} \zeta_3 \zeta_2
\brk
+\frac{275
            \zeta_2}{9}+\frac{3488 \zeta_3}{81}+\frac{152
            \zeta_4}{9}+\frac{224 \zeta_5}{9}-\frac{42727}{486}
\Biggr)
\brk
+ C_a n_f^2 \Biggl( -\frac{136 \zeta_2}{27}-\frac{560
            \zeta_3}{243}-\frac{44 \zeta_4}{27}-\frac{256}{6561} \Biggr)
\end{align}


\noindent\textbf{Discussion.} The explicit results for the rapidity
anomalous dimension in \eqn{eq:8} can be rewritten in a remarkable form:
\begin{align}
  \label{eq:11}
    \ugr_0 = & \, \gsoft_0
\nbrk
\ugr_1 = & \, \gsoft_1 - \beta_0 c^s_1 
\nbrk
\ugr_2 = & \, \gsoft_2 - 2 \beta_0 c^s_2 - \beta_1 c^s_1 + 2 C_a C_A \beta_0 \zeta_4
\end{align}
Eq.~(\ref{eq:11}) is interesting because it connects between very
different objects: the rapidity anomalous dimension $\gr$, the soft anomalous
dimension $\gamma_s$, the threshold constant $c_s$, and the QCD beta
function. Similar relation also holds in $\Neqfour$ SYM by dropping
the beta function terms in \eqn{eq:11}.

In the CSS formalism, the resummation of large $\qt$ logarithms is
controlled by two anomalous dimension, $A\bals{\mu}=\sum_{i=1} \as^i A_i$ and
$B\bals{\mu}=\sum_{i=1} \as^i B_i$. It is straightforward to express
these anomalous dimension in terms of the anomalous dimension in SCET,
see e.g. Ref.~\cite{1007.4005,lnssz}. In particular, we obtain the $B$ anomalous
dimension in the original CSS scheme through to three loops: 
\begin{align}
  \label{eq:12}
    B_1 = & \, \gamma^{\sss V}_0 - \ugr_0
\nbrk
  B_2 = & \, \gamma^{\sss V}_1 - \ugr_1 + \beta_0 c^{\sss V}_1
\nbrk
  B_3 = & \, \gamma^{\sss V}_2 - \ugr_2 + \beta_1 c^{\sss V}_1 + 2 \beta_0 \Big(
  c^{\sss V}_2 - \frac{1}{2} \big(c^{\sss V}_1\big)^2 \Big)
\end{align}
where $\gamma_{\sss V}$ is the anomalous dimension of hard function
results from matching QCD onto SCET. $c_{\sss V}$ is the scale
independent terms of the hard matching. For Drell-Yan production they
can be extracted from quark form factor~\cite{0902.3519,1001.2887,1004.3653}, while for Higgs
production from gluon form factor~\cite{0902.3519,1001.2887,1004.3653}, and additionally from effective coupling
of the Higgs boson to gluons~\cite{hep-ph/9708255}.
Eq.~(\ref{eq:12}) partially explains the close connection between
$\gr$ and $\gs$, because the combination $\gamma_{\sss V} - \gs$ is
given by the $\delta(1-x)$ part of the single pole in the QCD
splitting function~\cite{hep-ph/0403192}.
Substituting the actual numbers in \eqn{eq:12}, we find
\begin{gather}
  \label{eq:13}
  B^{\sss DY}_1 = -8, \quad B^{\sss DY}_2 = 13.3447 + 3.4138 \, n_f,
\nonumber
\\
B^{\sss DY}_3 = 7358.86 - 721.516 \, n_f + 20.5951 \, n_f^2
\end{gather}
for Drell-Yan production. For Higgs production, the results are
\begin{gather}
  \label{eq:14}
  B^{\sss H}_1 = -22 + 1.33333 \, n_f, \quad B^{\sss H}_2 = 658.881 -
  45.9712 n_f,
\nonumber
\\
B^{\sss H}_3 = 35134.6 - 7311.10 \,n_f + 293.017\, n_f^2
\nonumber
\\
- \big(836 + 184 \,n_f - 14.2222\, n_f^2\big) \ln \frac{m^2_t}{m^2_{\sss H}}
\end{gather}
The one and two-loop results are known for a long
time~\cite{Davies:1984hs,NUPHA.B256.413,hep-ph/0008152}. The three-loop results are
new. We note that numerically $B^{\sss DY}_3$ is quite large for $n_f
= 5$. 

In summary, we have presented the first calculation of soft function for
transverse-momentum resummation in rapidity RG formalism through to
three loops, using the rapidity regulator recently introduced in
Ref.~\cite{Li:2016axz}. As a by product, we have also obtained the fully
differential soft function to the same order.
Our calculation combine the use of bootstrap
technique and supersymmetric decomposition in transcendental weight. We found a
surprising relation between the anomalous dimensions for
the transverse-momentum resummation and the threshold resummation, whose
explanation calls for further investigation. Our three-loop results
pave the way for transverse-momentum resummation for production of
color neutral system at hadron colliders at N$^3$LL + NNLO
accuracy. The method and results of our calculation also make generalizing
$\qt$-subtraction method~\cite{hep-ph/0703012} to N$^3$LO promising.

\begin{acknowledgments}
We are grateful to useful conversation with Duff Neill, and helpful
comments on the manuscript by Iain Stewart. We thank Markus Ebert for pointing out typos in the supplemental material in earlier versions.  This work was supported by the Office of Nuclear Physics of the U.S. Department of Energy under Contract DE-SC0011090. Fermilab is operated by Fermi
Research Alliance, LLC under contract No.~DE-AC02-07CH11359 with the
U.S. Department of Energy.
\end{acknowledgments}

\appendix
\begin{widetext}

\section{One-loop beam function for $\qt$ resummation}
\label{sec:one-loop-soft}
The TMD beam function appearing in the factorization
formula in eq.~(2) using the exponential regulator of
Ref.~\cite{Li:2016axz}
differs from the corresponding beam function using the $\eta$
regulator~\cite{1202.0814}. The explicit expression for the TMD beam function with exponential regulator through to
NNLO can be extracted from the TMD parton
distribution functions, which are also known at
NNLO~\cite{Gehrmann:2014yya}. The idea is that the convolution of the
two beam function and the soft function in eq.~(2) is independent
of rapidity regulator and therefore is identical to the convolution of
two transverse-momentum dependent parton
distribution functions of Ref.~\cite{Gehrmann:2014yya}. The
TMD beam function can also be computed directly using
the exponential regulator in Ref.~\cite{Li:2016axz}. The two approaches
give identical result as they should. The details of the direct
calculation for the beam function using exponential regulator will be
given elsewhere. For the reader's convenience, we give below their
explicit expressions for Drell-Yan production through to NLO. At
the perturbative scale, the renormalized beam function can written as the convolution of
coefficient function and the usual parton distribution functions:
\begin{align}
  \label{eq:16}
\tag{$1$S}
  B_{i/N}(z,L_b,\LQ) = \, & \sum_j \int^1_{z}
  \frac{d\xi}{\xi} \mathcal{I}_{ij}(\xi, L_b, \LQ) f_{j/N}(z/\xi,\mu)
+ \Ord( \vecb^{\, 2} \Lambda_{\sss \rm QCD}^2) \, 
\end{align}
where $L_b = \ln \vecb^{\, 2} \mu^2/b^2_0$ and $\LQ = \ln Q^2/\nu^2$.
At LO, the non-vanishing coefficient functions
are $\mathcal{I}_{0,qq}(z,L_b,\LQ)=\mathcal{I}_{0,\bar{q}\bar{q}}(z,L_b,\LQ) =
 \delta(1-z)$. At NLO we find
\begin{align}
  \label{eq:15}
\tag{$2$S}
 \mathcal{I}_{1,qq}(z,L_b,\LQ) = \,& -\frac{1}{2} \delta(1-z) \big[ \gcusp_0
                                     L_b \LQ + \gamma^r_0 \LQ
 + ( \gamma^s_0 +        \gamma^{\sss V}_0) L_b  \big] 
- P_{0,qq}(z) L_b   + 2C_F (1-z) \,,
\nbrk
\mathcal{I}_{1,qg}(z,L_b,\LQ) = \, & 4 T_{F} z (  1 - z
                                     ) - 
                                     P_{0,qg}(z) L_b
\tag{$3$S}
\end{align}
where for Drell-Yan production $\Gamma^{\rm cusp,\sss DY}_0 = 4C_F$, $\gamma^r_0 =
\gamma^r_s = 0$, $\gamma^{\sss V,DY}_0 = - 6 C_F $, and $P_{0,ij}(z)$
are the usual LO splitting function
\begin{align}
  \label{eq:17}
\tag{$4$S}
  P_{0,qq}(z) =\,& 3 C_F \delta(1-z) + 4C_F\left[\frac{1}{1-z}
                   \right]_+ -2 C_F (1+z)
\nbrk
P_{0,qg}(z) = \, & 2 T_F (1 - 2 z + 2 z^2) 
\tag{$5$S}
\end{align}
where $T_F = 1/2$ for QCD. The remaining coefficient functions for
Drell-Yan production at NLO can be obtained by charge conjugation. We
note that the coefficient functions have the interesting property
that in their $\delta(1-z)$ terms only scale dependent pieces
exist. All the constant terms reside in the soft function. To the best
of our knowledge this is a unique feature of our
rapidity regulator. 

\section{The fully differential soft function through to three loops
  including scale dependent terms}
\label{sec:scale-depend-soft}

The RG equation eq.~(6) for the fully differential soft function can be solved
to all orders up to scale independent terms. Through to three loops it reads
\begin{align}
  \label{eq:18}
  \sfd(\vecb,\mu,\nu) = \, & \exp\Bigg\{ \as \Bigg[\esfd_1\big|_{\mu=\nu}+\gsoft_0
              L_\nu+\frac{\gcusp_0 L_\nu^2}{2} \Bigg] 
\nbrk
\,& + \as\Bigg[ L_{\nu } \left(\beta _0
    \big(\esfd_1|_{\mu=\nu}\big)+\gsoft_1\right)+\big(\esfd_2|_{\mu=\nu}\big)+L_{\nu
    }^2 \left(\frac{\gcusp_1}{2}+\frac{\beta _0
    \gsoft_0}{2}\right)+\frac{1}{6} \beta _0 \gcusp_0 L_{\nu }^3
\Bigg]
\brk
+ \as^3 \Bigg[
L_{\nu }^2 \left(\beta _0^2 \big(\esfd_1|_{\mu=\nu}\big)+\beta _0
    \gsoft_1+\frac{\beta _1
    \gsoft_0}{2}+\frac{\gcusp_2}{2}\right)+L_{\nu } \left(2 \beta _0
    \big(\esfd_2|_{\mu=\nu}\big)+\beta _1
    \big(\esfd_1|_{\mu=\nu}\big)+\gsoft_2\right)
\brk
+\big(\esfd_3|_{\mu=\nu}\big)+L_{\nu }^3 \left(\frac{1}{3} \beta _0^2 \gsoft_0+\frac{\beta _0 \gcusp_1}{3}+\frac{\beta _1 \gcusp_0}{6}\right)+\frac{1}{12} \beta _0^2 \gcusp_0 L_{\nu }^4
\Bigg] + \Ord(\as^4) \Bigg\}
\tag{$6$S}
\end{align}
where $L_\nu = \ln(\nu^2/\mu^2)$.
The one and two-loop constants $\esfd_{1,2}|_{\mu=\nu}$ are
first computed in Ref.~\cite{1105.5171} and reproduced in Ref.~\cite{Li:2016axz}:
\begin{align}
  \label{eq:19}
\tag{$7$S}
  \esfd_1|_{\mu=\nu}= \, & 4 C_a H_{2} + c^s_1
\nbrk
\esfd_2 |_{\mu=\nu} = \, & C_A C_a \left(-8 \zeta_2
                           H_{2}+\frac{268}{9} H_{2}+\frac{44}{3}
                           H_{3}-8 H_{4}-\frac{44}{3} H_{2,1}-8
                           H_{2,2}-16 H_{3,1}-16 H_{2,1,1}\right)
\brk
+C_a n_f \left(-\frac{40}{9} H_{2}-\frac{8}{3} H_{3}+\frac{8}{3}
                           H_{2,1}\right) + c^s_2
\tag{$8$S}
\end{align}
The three-loop constant is given in eq.~(8), and $c^s_1$ and
$c^s_2$ can be found in \eqn{eq:c3}. The argument of the
HPLs is $x = - \vecb^{\, 2} \nu^2/b^2_0$.
From the result in \eqn{eq:18}, we can derive the TMD soft function
by taking the limit of $1/\nu^2 \to 0$, and keeping only the
non-vanishing terms. From there we can extract the rapidity anomalous
dimension as well as the constant terms of the soft function. The
result for $S_\perp$ through to three loops reads:
\begin{align}
  \label{eq:20}
  S_\perp(\vecb,\mu,\nu) = \, & \exp\Bigg\{
\as \Bigg[ 
c^{\perp}_1+\frac{1}{2} \gcusp_0 L_b^2+\ugr_0 L_r-L_b \left(\gsoft_0+\gcusp_0 L_r\right)
\Bigg] 
\brk
+
\as^2 \Bigg[
c^{\perp}_2+\ugr_1 L_r+\frac{1}{6} \gcusp_0 L_b^3 \beta _0+L_b^2
                                \left(\frac{\gcusp_1}{2}-\frac{\gsoft_0
                                \beta _0}{2}-\frac{1}{2} \gcusp_0 L_r
                                \beta _0\right)
\brk
+L_b \left(-\gsoft_1+c^{\perp}_1 \beta _0+L_r \left(-\gcusp_1+\ugr_0 \beta _0\right)\right)
\Bigg]
\brk
+
\as^3 \Bigg[
c^{\perp}_3+\ugr_2 L_r+\frac{1}{12} \gcusp_0 L_b^4 \beta _0^2+L_b^3
                                \left(\frac{\gcusp_1 \beta
                                _0}{3}-\frac{1}{3} \gsoft_0 \beta
                                _0^2-\frac{1}{3} \gcusp_0 L_r \beta
                                _0^2+\frac{\gcusp_0 \beta
                                _1}{6}\right)
\brk
+L_b^2 \left(\frac{\gcusp_2}{2}-\gsoft_1 \beta _0+c^{\perp}_1 \beta _0^2-\frac{\gsoft_0 \beta _1}{2}+L_r \left(-\gcusp_1 \beta _0+\ugr_0 \beta _0^2-\frac{\gcusp_0 \beta _1}{2}\right)\right)
\brk
+L_b \left(-\gsoft_2+2 c^{\perp}_2 \beta _0+c^{\perp}_1 \beta _1+L_r
                                \left(-\gcusp_2+2 \ugr_1 \beta
                                _0+\ugr_0 \beta _1\right)\right)
\Bigg] 
+ \Ord(\as^4) \Bigg\}
\tag{$9$S}
\end{align}
where $L_r = \ln \big(\nu^2 \vecb^{\, 2}/b^2_0\big)$ is the rapidity
logarithm, $L_b = \ln (\vecb^{\, 2}\mu^2/b^2_0)$, and the scale
independent constant at one and two loop(s) are~\cite{Li:2016axz}:
\begin{align}
  \label{eq:21}
\tag{$10$S}
  c^{\perp}_1 = \, & -2 C_a \zeta_2
\nbrk
c^{\perp}_2 = \, & C_A C_a \left(-\frac{67 \zeta_2}{3}-\frac{154 \zeta_3}{9}+10 \zeta_4+\frac{2428}{81}\right)+C_a n_f \left(\frac{10 \zeta_2}{3}+\frac{28 \zeta_3}{9}-\frac{328}{81}\right)
\tag{$11$S}
\end{align}
The three loop expression is given in eq.~(10).
It is straightforward to check that $S_\perp(\vecb, \mu, \nu)$
satisfies both the usual RG equation in eq.~(3) and rapidity RG
equation in eq.~(4). Note that in eq.~(4), the rapidity
anomalous dimension is evaluated at the scale $\mu = b_0/|\vecb|$. 

For fixed $\vecb$, The fully differential soft function $\sfd$
interpolate between TMD soft function at $1/\nu^2 \to 0$, and
threshold soft function at $\nu^2 \to 0$. This is illustrated
numerically in Fig.~\ref{fig:a} at three different orders in $\as$ by
varying $\nu^2$ while keeping $\mu^2 = b^2_0/\vecb^{\, 2}$ fixed.
\begin{figure}[!ht]
  \begin{subfigure}{0.3\linewidth}
    \centering
    \includegraphics[width=1\linewidth]{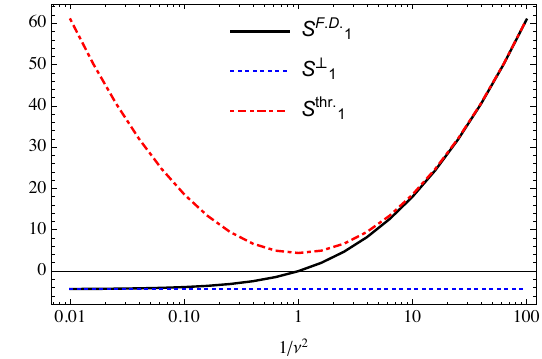}
  \end{subfigure}
  \begin{subfigure}{0.3\linewidth}
    \centering
    \includegraphics[width=1\linewidth]{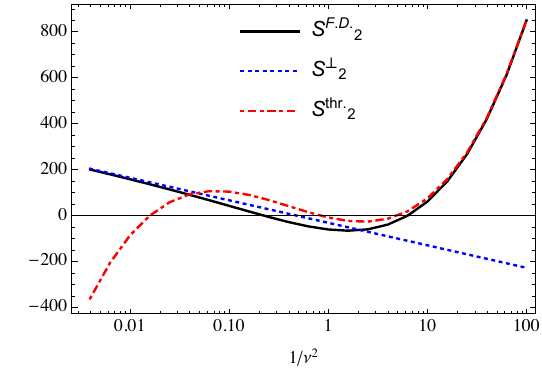}
  \end{subfigure}
  \begin{subfigure}{0.3\linewidth}
    \centering
    \includegraphics[width=1\linewidth]{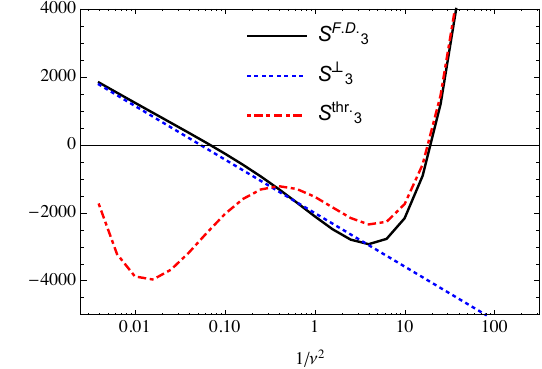}
  \end{subfigure}
  \caption{\raggedright Illustration of the asymptotic behavior of $\sfd$ through
    to three loops as a
    function of $\nu$, with $\mu^2 = b^2_0/\vecb^{\, 2}$. Depicted are
    the coefficients of $\as^n$ at successive order, with the
    numerical value of color factor for Drell-Yan process substituted
    in. It can be seen that at asymptotically large $\nu$ ($1/\nu^2 \to
    0$), the fully differential soft function $\sfd$~(the black solid line) approaches the $\qt$
  soft function $S_\perp$~(the blue dotted line). While at small
  $\nu$~($1/\nu^2 \to \infty$), it approaches the threshold soft
  function $S_{\rm thr.}$~(the red dot-dashed line). At $\Ord(\as)$ the
  $\qt$ soft function is a horizontal line because $\ugr_0 = 0$ at
  this order, see the first line of eq.~(9). }
  \label{fig:a}
\end{figure}

\section{Anomalous dimensions and Wilson coefficients}
\label{sec:anomalous-dimension}
In this appendix we summarize the relevant anomalous dimensions and Wilson
coefficients. The QCD beta functions through to two loops are:
\begin{align}
  \label{eq:10}
\tag{$12$S}
  \beta_0 = &\, \frac{11 C_A}{3}-\frac{2 n_f}{3}
\nbrk
\beta_1 = &\, \frac{34 C_A^2}{3}-\frac{10 C_A n_f}{3}-2 C_F n_f
\tag{$13$S}
\end{align}

The QCD cusp anomalous dimension through to three loops have been
computed in Ref.~\cite{hep-ph/0403192}. The results are
\begin{align}
  \gcusp_{0} = & \, 4 C_a
\nbrk
  \gcusp_{1} = & \, C_A C_a \left(\frac{268}{9}-8 \zeta_2\right)-\frac{40 C_a n_f}{9}
\nbrk
  \gcusp_{2} = & \, C_A^2 C_a \left(-\frac{1072 \zeta_2}{9}+\frac{88
      \zeta_3}{3}+88 \zeta_4+\frac{490}{3}\right)
+C_A C_a n_f \left(\frac{160 \zeta_2}{9}-\frac{112
    \zeta_3}{3}-\frac{836}{27}\right)
\brk
+C_a C_F n_f \left(32 \zeta_3-\frac{110}{3}\right)-\frac{16 C_a
                 n_f^2}{27}
\label{eq:c1}
\tag{$14$S}
\end{align}
The threshold soft anomalous dimensions are~\cite{1412.2771}
\begin{align}
    \gsoft_{0} = & \, 0
\nbrk
  \gsoft_{1} = & \, C_A C_a \left(\frac{22 \zeta_2}{3}+28 \zeta_3-\frac{808}{27}\right)+C_a n_f \left(\frac{112}{27}-\frac{4 \zeta_2}{3}\right)
\nbrk
  \gsoft_{2} = & \, C_A^2 C_a \left(-\frac{176}{3} \zeta_3
    \zeta_2+\frac{12650 \zeta_2}{81}+\frac{1316 \zeta_3}{3}-176
    \zeta_4
-192 \zeta_5-\frac{136781}{729}\right)
\brk
+C_A C_a n_f \left(-\frac{2828
  \zeta_2}{81}-\frac{728 \zeta_3}{27}
+48 \zeta_4+\frac{11842}{729}\right)+C_a C_F n_f \left(-4
\zeta_2-\frac{304 \zeta_3}{9}
-16 \zeta_4+\frac{1711}{27}\right)
\brk
+C_a n_f^2 \left(\frac{40 \zeta_2}{27}-\frac{112
                 \zeta_3}{27}+\frac{2080}{729}\right)
\label{eq:c2}
\tag{$15$S}
\end{align}
The constants of the threshold soft function are~\cite{1412.2771}
\begin{align}
  c^s_{1} = & \, 2 C_a \zeta_2
\nbrk
  c^s_{2} = & \, C_A C_a \left(\frac{67 \zeta_2}{9}-\frac{22
      \zeta_3}{9}-30 \zeta_4+\frac{2428}{81}\right)
+C_a n_f \left(-\frac{10 \zeta_2}{9}+\frac{4 \zeta_3}{9}-\frac{328}{81}\right)
\nbrk
  c^s_{3} = & \, C_A^2 C_a \left(\frac{1072
      \zeta_3^2}{9}-\frac{220}{9} \zeta_2 \zeta_3-\frac{87052
      \zeta_3}{243}-\frac{20371 \zeta_2}{729}
-\frac{9527 \zeta_4}{27}-\frac{968 \zeta_5}{9}+\frac{8506
  \zeta_6}{27}+\frac{5211949}{13122}\right)
\brk
+C_A C_a n_f \left(-\frac{8}{9} \zeta_3 \zeta_2+\frac{2638
    \zeta_2}{729}+\frac{1216 \zeta_3}{81}+\frac{928 \zeta_4}{27}
-\frac{16 \zeta_5}{3}-\frac{412765}{6561}\right)+C_a C_F n_f
\left(\frac{16}{3} \zeta_3 \zeta_2-\frac{55 \zeta_2}{9}
\right. \nn \\ & \left.
+\frac{2840 \zeta_3}{81}+\frac{152 \zeta_4}{9}+\frac{224
  \zeta_5}{9}-\frac{42727}{486}\right)
+C_a n_f^2 \left(-\frac{8 \zeta_2}{81}+\frac{880
                 \zeta_3}{243}+\frac{52
                 \zeta_4}{27}-\frac{256}{6561}\right)
\label{eq:c3}
\tag{$16$S}
\end{align}
Note that eqs.~(\ref{eq:c1}), (\ref{eq:c2}) and (\ref{eq:c3}) obey Casimir scaling and therefore
is process independent.  The hard functions are process
dependent and can be extracted from quark and gluon
form factors~\cite{0902.3519,1001.2887,1004.3653}, and additionally from effective coupling
of the Higgs boson to gluons~\cite{hep-ph/9708255}. They were needed in
connecting the rapidity anomalous dimension to the $B$ coefficients in
 eq.~(12).
The hard anomalous dimensions  for Drell-Yan production are
\begin{align}
  \gamma^{\sss V,DY}_0 = & \, -6 C_F
\nbrk 
  \gamma^{\sss V,DY}_1 = & \, C_A C_F \left(-22 \zeta_2+52
    \zeta_3-\frac{961}{27}\right)
+C_F^2 (24 \zeta_2-48 \zeta_3-3)+C_F n_f \left(4 \zeta_2+\frac{130}{27}\right)
\tag{$17$S}
\end{align}
For Higgs production, they are
\begin{align}
  \gamma^{{\sss V},h}_0 = & \, -2 \beta_0
\nbrk 
  \gamma^{{\sss V},h}_1 = & \,C_A^2 \left(\frac{22 \zeta_2}{3}+4 \zeta_3-\frac{1384}{27}\right)+C_A n_f \left(\frac{256}{27}-\frac{4 \zeta_2}{3}\right)+4 C_F n_f
\tag{$18$S}
\end{align}
The constants of hard function for Drell-Yan production are
\begin{align}
  c^{\sss V, DY}_1 = & \, C_F (14 \zeta_2-16)
\nbrk 
  c^{\sss V,DY}_2 = & \, C_A C_F \left(\frac{1061 \zeta_2}{9}+\frac{626
      \zeta_3}{9}-16 \zeta_4-\frac{51157}{324}\right)
+C_F^2 \left(-166 \zeta_2-60 \zeta_3+201 \zeta_4+\frac{511}{4}\right)
\brk
+C_F n_f \left(-\frac{182 \zeta_2}{9}+\frac{4 \zeta_3}{9}+\frac{4085}{162}\right)
\tag{$19$S}
\end{align}
And for Higgs production they are
\begin{align}
  c^{{\sss V},h}_1 = & \, C_A (14 \zeta_2+10)-6 C_F
\nbrk 
  c^{{\sss V},h}_2 = & \, C_A^2 \left(14 \ln \frac{m^2_{\sss
                       H}}{m^2_{t}}+\frac{755
                       \zeta_2}{3}-\frac{286 \zeta_3}{9}+185
                       \zeta_4+\frac{23827}{162}\right)+C_A C_F
                       \left(-22 \ln \frac{m^2_{\sss
                       H}}{m^2_t}-84
                       \zeta_2-\frac{290}{3}\right)
\brk
+C_A n_f
                       \left(-\frac{50 \zeta_2}{3}-\frac{92
                       \zeta_3}{9}-\frac{2255}{81}\right)-\frac{5
                       C_A}{6}+36 C_F^2+C_F n_f \left(8 \ln
                       \frac{m^2_{\sss H}}{m^2_t}+16
                       \zeta_3-\frac{82}{3}\right)-\frac{4 C_F}{3}
\label{eq:c}
\tag{$20$S}
\end{align}
 where we have set the matching scale in the Higgs effective
 theory to be $\mu = m_{\sss H}$. Note that \eqn{eq:c} comes from the
 product of Higgs effective theory Wilson coefficient and gluon form
 factor expanded to the given order.


\end{widetext}

\end{document}